# Main Manuscript for

# Alterations of electrocortical activity during hand movements induced by motor cortex glioma

Yihan Wu[1,6,†], Tao Chang[2,†], Siliang Chen[2], Xiaodong Niu[2], Yu Li[3], Yuan Fang[2], Lei Yang[3], Yixuan Zong[2], Yaoxin Yang[3], Yuehua Li[4], Mengsong Wang[4], Wen Yang[4], Yixuan Wu[3], Chen Fu[3], Xia Fang[5], Yuxin Quan[2], Xilin Peng[3], Qiang Sun[7], Marc M. Van Hulle[7], Yanhui Liu[2], Ning Jiang[1,6], Dario Farina[8*], Yuan Yang [2*], Jiayuan He[1,6*], and Qing Mao[2*]

**1**. National Clinical Research Center for Geriatric, West China Hospital of Sichuan University, Chengdu, Sichuan 610017, China.

**2**. Department of Neurosurgery, West China Hospital of Sichuan University, No. 37 Guo Xue Xiang, Chengdu 610041, China.

**3**. Department of Anesthesiology, West China Hospital of Sichuan University, Chengdu, 610041, Sichuan, P.R. China.

**4**. Operating Room, Department of Anesthesiology, West China Hospital of Sichuan University, Chengdu, Sichuan, China.

**5**. College of Mechanical Engineering, Sichuan University, Chengdu, Sichuan 610065, China.

**6**. Med-X Center for Manufacturing, Sichuan University, Chengdu, Sichuan 610017, China.

**7**. Laboratory for Neuro- and Psychophysiology, Department of Neurosciences, KU Leuven, Leuven, Belgium.

**8**. Department of Bioengineering, Imperial College of Science, Technology and Medicine, London, UK.

† Yihan Wu, and Tao Chang contributed equally to this work, and should be considered as co-first authors.

* Correspondence: Dario Farina, Yuan Yang, Jiayuan He, and Qing Mao

**Email:**  d.farina@imperial.ac.uk (DF); yangyuan@wchscu.cn (YY); jiayuan.he@wchscu.cn (JYH); qingmao2000@163.com (QM)

This PDF file includes:

>   Main Text
>   Figures 1 to 4

**Abstract**

Glioma cells can reshape functional neuronal networks by hijacking neuronal synapses, leading to partial or complete neurological dysfunction. These mechanisms have been previously explored for language functions. However, the impact of glioma on sensorimotor functions is still unknown. Therefore, we recruited a control group of patients with unaffected motor cortex and a group of patients with glioma-infiltrated motor cortex, and recorded high-density electrocortical signals during finger movement tasks. The results showed that glioma suppresses task-related synchronization in the high-gamma band and reduces the power across all frequency bands. The resulting atypical motor information transmission model with discrete signaling pathways and delayed responses disrupts the stability of neuronal encoding patterns for finger movement kinematics across various temporal-spatial scales. These findings demonstrate that gliomas functionally invade neural circuits within the motor cortex. This result advances our understanding of motor function processing in chronic disease states, which is important to advance the surgical strategies and neurorehabilitation approaches for patients with malignant gliomas.

**Significance Statement**

The research fills a gap in understanding the impact of motor cortex gliomas on fine motor control, particularly intricate finger movements. By employing electrocorticography on glioma patients, the study documents chaotic electrophysiological activities in the affected motor cortex, disrupting standard motor information transmission and compromising the consistency of neural encoding patterns across different spatiotemporal dimensions. These novel insights contribute to a deeper comprehension of motor function reorganization post-glioma and may inform refined surgical interventions and neurorehabilitation strategies for patients with these tumors in speech and motor areas.

**Main Text**

**Introduction**

Glioma, characterized by infiltrative growth, is the most common primary intracranial tumor of the central nervous system in adults (1). It disrupts and damages the information transmission of the central nervous system. Normal functions are therefore affected depending on the location of the glioma. Motor function is the core objective of the nervous system in humans and other animals. For motor-related cortex glioma, the risk of permanent motor dysfunction ranges from 1.5% to 24.0% (2-6). A better understanding of the biology of motor functional network within glioma-infiltrated motor cortex is needed for understanding its impact on motor functions, and critically, for minimizing the risks of neurologic sequelae (7, 8).
Previous research has demonstrated detailed interactions between neurons and gliomas, occurring at both the cellular and network levels. In this dynamic interplay, neuronal activity has been identified as a driver for glioma progression. On the other hand, gliomas have been found to enhance neuronal excitability, partially participating in neural signal synchronization and processing under chronic pathology (9, 10). This indicates that the cortical neuronal activities underlying motor functions could be altered following glioma infiltration. The alterations are related to the glioma-neuron interaction, and can provide insights into the structural reorganization and functional plasticity deriving from glioma cells in the motor cortex (11). High-density electrocorticography (ECoG) provides millimeter-scale spatial resolution and millisecond-scale temporal resolution. These characteristics enable the accurate identification of functional anatomical connections and timing of neurophysiological mechanisms at the circuit level (12-14). For instance, ECoG was used to identify alterations in the neuronal power spectrum with tumor-infiltrated glioma during speech modulation (15-17). However, limited research has delved into the impact of glioma on the neuronal processing of the motor cortex, especially during movements of the fingers – the most dexterous task of humans, associated to a large area in the motor cortex. Consequently, the specific alterations in the neuronal physiology, the underlying mechanism governing information transfer, and the motor encoding in the brain in the presence of glioma remain unclear.

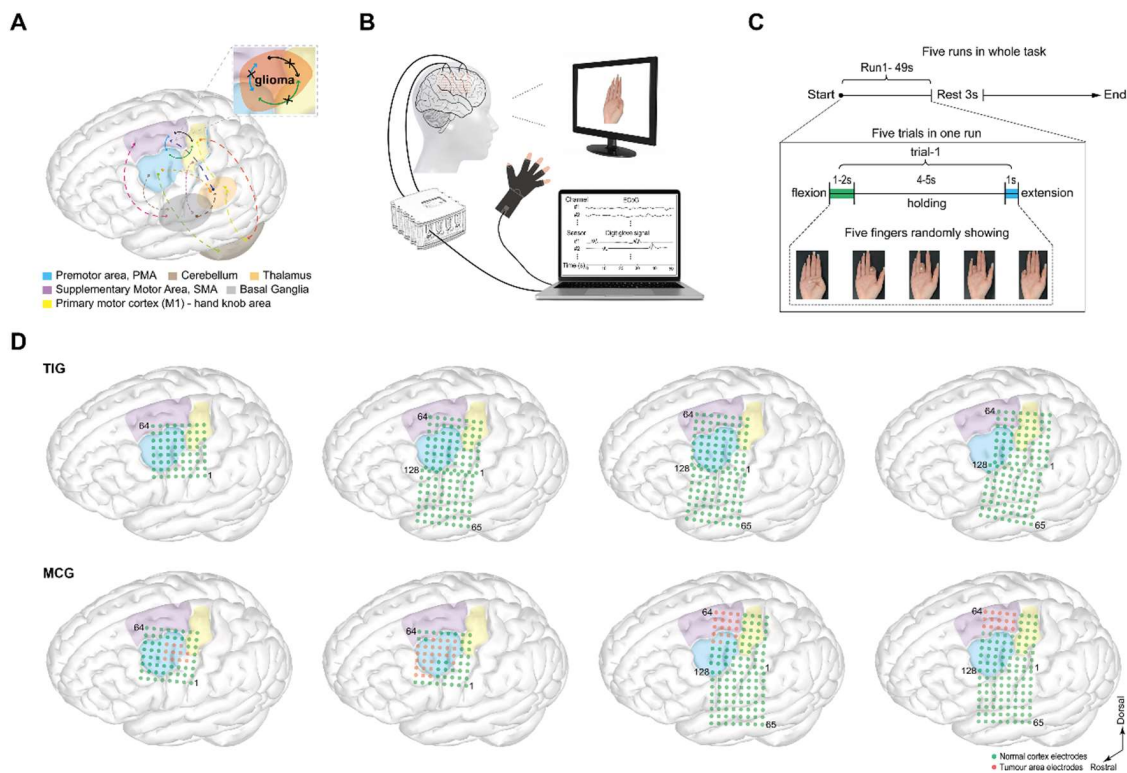

**Fig.1 Motor circuits, experimental workflow, and electrode localization. (A)**, The topographic projections from these specific areas are represented by colored arrows. In cases where a glioma occurs in the motor cortex, such as the SMA, PMA, and M1, we hypothesized the motor cortex around the glioma may represent abnormal electrophysiological activity, impede the transmission of motor information, and develop an atypical encoding model for finger kinematics across various brain regions and rhythms. **(B)**, Schematic of experimental workflow. In participants with motor cortex gliomas, we applied high-density ECoG to investigate the neuronal response alterations during finger movement. The invasive electrophysiologic signals were recorded and processed for analysis. **(C)**, Experimental paradigm. The experiment comprised five sequential runs, each lasting for about 49 seconds, and interspersed with 3-second intervals of rest between runs, resulting in a total duration of approximately 5 minutes. **(D)**, Electrode localization. On the MNI template, 64-128 electrodes were positioned over the cortical regions exposed in each participant.

This work investigated neuronal responses of the human brain, comparing cases with and without glioma-infiltrated motor cortex during finger movement tasks. The alteration of neural oscillations from glioma was studied from three perspectives: the time and frequency domain content of cortical electrical activities, the information flow in motor neuronal circuits, and the decoding performance among different regions in the motor cortex. We found that the cortical information transmission and information encoding in glioma-infiltrated areas are reduced and that more neural tissue surrounding the tumor is activated during motion. These findings underline that the motor cortex around the glioma may exhibit abnormal electrophysiological activity, developing an atypical model for motor information transmission. This, in turn, disrupt the stability of neuronal encoding patterns for finger movement kinematics, across various temporal-spatial scales (Fig.1A). These insights advance our understanding of the underlying mechanisms of motor functional reconfiguration, integrate cerebral structure and function plasticity into the surgical strategies, and inform effective neurorehabilitation approaches for patients with glioma involving eloquent cortex.

## Results

### Participant characteristics.
The cohort consisted of eight participants, categorized into two groups: four individuals with temporo-insular glioma (TIG) served as the control group (healthy motor cortex), where the tumor did not invade the motor regions or motor transmission pathways; and four patients with motor cortex glioma (MCG), where the tumor infiltrated the motor cortex, were classified as the case group (Tab. S1). According to the 2021 WHO classification (18), all gliomas were diagnosed as grade 2-3, encompassing oligodendroglioma and astrocytomas. The lesions were located on the left hemisphere side in all predominantly right-handed participants. Importantly, none of the participants presented any notable clinical symptoms. Both groups exhibited similar clinical characteristics, with no statistically significant differences in the distribution of pathological features or preoperative functional evaluation scores ($p > 0.05$) (Tab. S2, Fig. S1-2).

### Motor cortex glioma and electrophysiological activity.
The first analysis focused on individual electrode performance to assess local neural responses. Specifically, for each participant, the electrode in the M1 hand area with the most pronounced power alteration during movement execution was selected for subsequent detailed analysis (Fig. 2A). Temporal-spectral variation was conducted by calculating Event-Related Spectral Perturbation (ERSP) in the range of 4 to 140 Hz. In comparison to the TIG group, the MCG group exhibited an extended time course of power decrease, i.e., Event-Related Desynchronization (ERD) in the low-frequency band (< 50Hz). The TIG group showed a short-lived power increase, i.e., Event-Related Synchronization (ERS), in the high-frequency band, which was absent in the MCG group (Fig. 2B). Furthermore, there were strong temporal correlations of ERSP curves between the two groups in the low-frequency bands, including alpha and beta. Comparing the time-series curves of ERSP between the two groups in the Gamma-3 and Gamma-4 bands, these two ERSP curves displayed differences at multiple time points ($p<0.05$), indicating that the disparity in neuronal activity between the two groups became more pronounced at higher frequencies. Meanwhile, the correlation between the average ERSP curves decreased as the frequency increased (Fig. 2C). These results suggest that the impact of glioma was mainly on the high-frequency band of the neuron electrical activities leading to the suppression of task-related ERS features. This observation was further confirmed by the distinct discrepancies observed in the high-gamma rhythm (60-140Hz) during the movement execution phase (Fig. 2D). Lastly, the abstract power of each sub-band was compared. The power distribution was similar between the two groups, peaking in the beta band and gradually decreasing with increasing frequency. A significant power reduction was observed across all bands in the MCG group (Fig. 2E). These findings demonstrated that the M1 cortex surrounding the glioma might still retain the ability to coordinate neuronal oscillations in response to the dynamics of individual fingers in the glioma-bearing motor cortex, but progressive disruption occurred starting from the high-gamma rhythm.

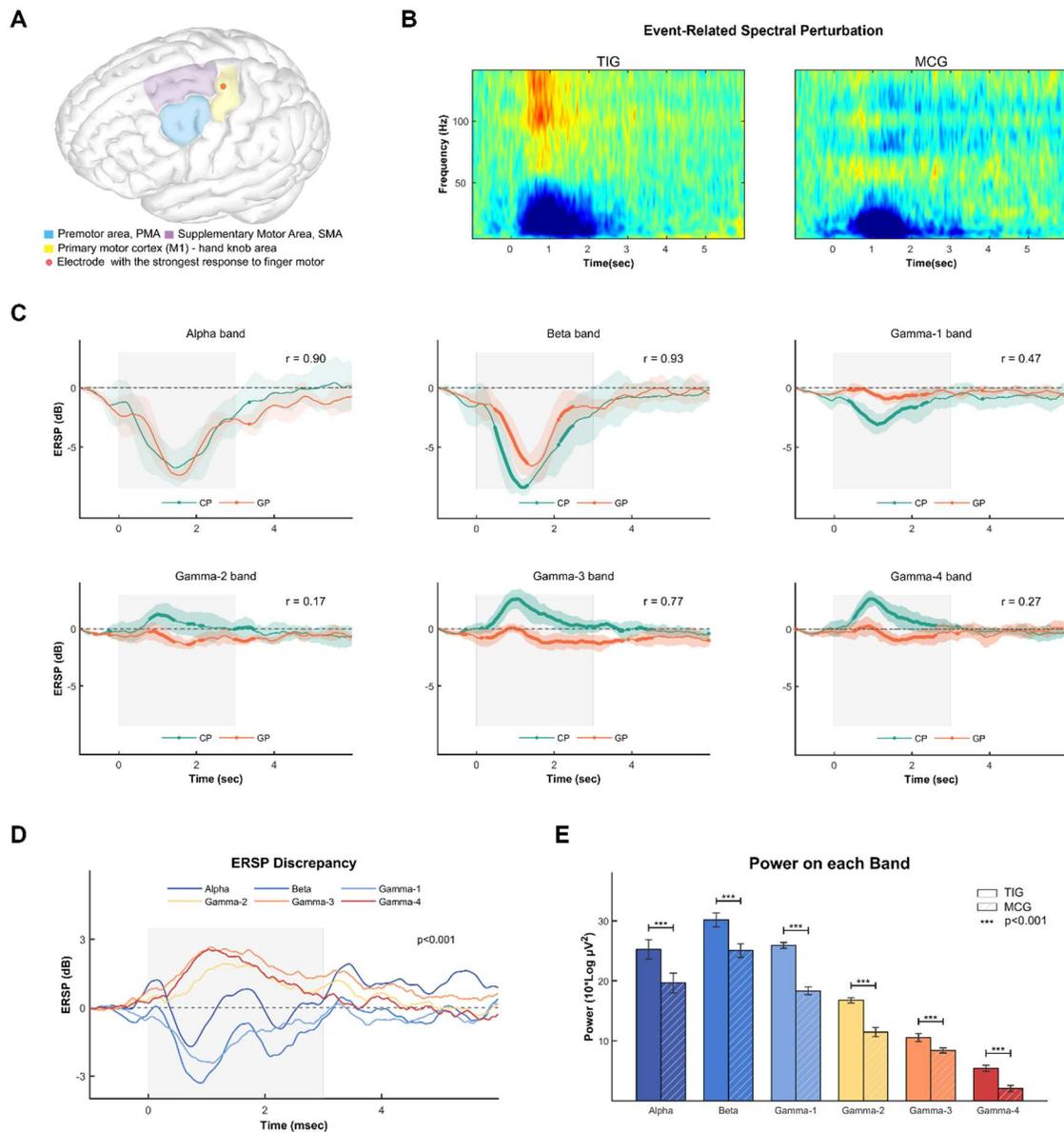

**Fig.2 Time-frequency domain alteration of neuronal activity in primary motor cortex. (A)**, Illustration of the electrode location in the M1 hand knob area. During the task, there is a distinct neuronal activation in the M1 area, responsible for controlling manipulative hand movement. **(B)**, Differences in the time-frequency domain between the two groups of patients. In the TIG group, Event-Related Desynchronization (ERD) and Event-Related Synchronization (ERS) are evident in the low and high-frequency, respectively. Conversely, the MCG group presents a diminished ERS and other disrupted features in the high-frequency band, along with an extended duration of ERD in the low-frequency band after the start of the task (at 0 seconds). **(C)**, Comparison of mean Event-Related Spectral Perturbation (ERSP) across various frequency bands between the two groups. As the frequency increased, the ERSP curves of both groups gradually desynchronize, and more time points in the high-frequency band show statistically significant differences. The time points with significant differences are highlighted in bold, and the Pearson correlation coefficient is calculated between the curves of the two groups. **(D)**, Discrepancies in ERSP between two groups in various bands. Prior to the task initiation, no significant differences are noted. However, following the task initiation, increased disparities are evident at each time point between the two ERSP curves in the high Gamma rhythm (60-140Hz), with this pattern gradually diminishing upon completion of motor task. **(E)**, The neuronal power across each subband between the two groups. A peak in the Beta rhythm is evident,

gradually decreasing as frequency increased within both groups, with a distinctive power reduction across each subband in MCG (p<0.001).

## Discrete signaling pathways and delayed response.

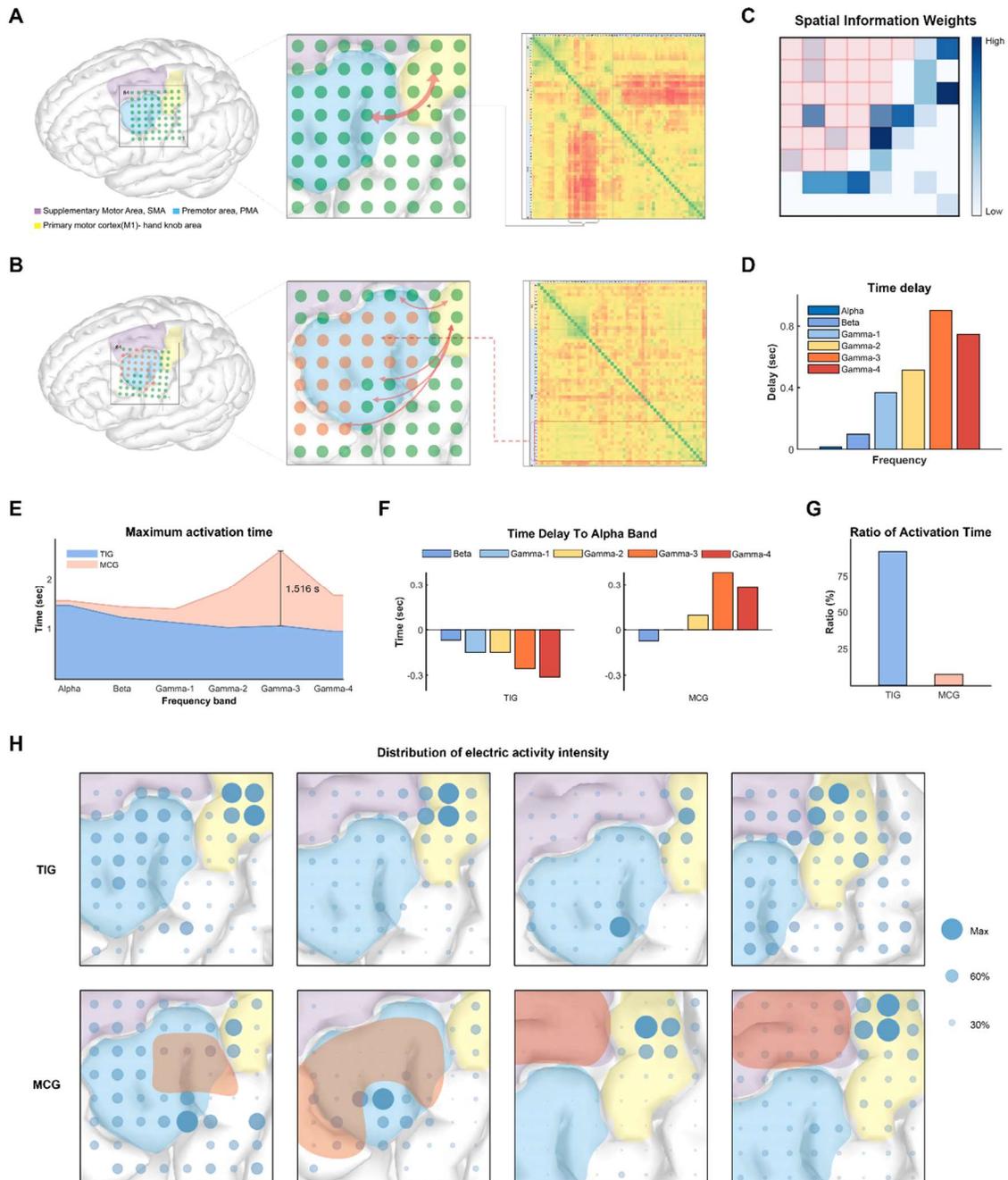

**Fig.3 Modification of motor information flow in the spatial and temporal domain. (A)**, The spatial-temporal power distribution model and information flow model in TIG 01 patient. The main activated areas are the M1 hand area and PMA. The bidirectional arrow symbolizes the exchange of motor information between the M1 and PMA. **(B)**, Atypical motor information transmission between the M1 hand area and regions around the lesion in MCG 02 patient. The areas of activation mainly involve the M1 hand area and the region surrounding the tumor. The red color indicates the extent of the lesion. The tumor region shows limited engagement in the neuronal response, while the cortex around the glioma displays a more prominent engagement in the process of motor information transmission. The bidirectional arrow represents the disrupted pathway of information flow between the M1 hand area and the cortex surrounding gliomas. **(C)**,

A weight diagram of electrodes for motor information flow in MCG 02 patient. **(D)** and **(E)**, The maximum activation time and the lag time difference of information transmission across various bands in the two groups. The time-lag index demonstrates a rising trend with increasing rhythm, culminating in a peak in the Gamma-3 band in the MCG group. Additionally, the impact of the tumor on information transmission is predominantly reflected in the high Gamma band. **(F)**, To eliminate the influence of individual differences for time-lag index across individuals, the statistical analysis involves examining the time-lag metric of each frequency band relative to its Alpha band for each individual. By comparing the differences in the corresponding frequency bands between the two groups, it ultimately leads to similar conclusions as D and E. **(G)**, During action execution in MCG group, there is a significant reduction in the ratio of ERS duration in the high-frequency band, indicating weaker activation intensity in activation time within this frequency range. **(H)**, The spatial distribution of average power in each individual. The tumor core region exhibits relatively lower average power, while abnormal high-power is observed in the cortex around glioma. The size of each circle is positively correlated with the power normalized value.

Except for the local neural activities, the interactions among brain regions might be affected by glioma as well. The bidirectional interaction transmission between the M1 hand area and PMA in TIG 01 patient was identified based on the classical motor theory principles, as illustrated in Fig. 1A (Fig. 3A). In the MCG group, this interaction manifested in an atypical manner with motor cortex glioma (Fig. 3B). For instance, the motor cortex with tumor infiltration was scarcely involved in the transmission of motor information, while the cortex surrounding the glioma demonstrated more prominent engagement during the task process (Fig. 3C). Further, it was observed that the power of neural activation in the tumor core region was relatively low, while abnormal high power was observed in the surrounding regions (Fig. 3H). These results suggest a distinctive pathways of information flow in the MCG patients.

To further access the efficiency of motor information transmission, we defined the maximum activation time as the time needed to reach the peak of the ERSP curve. An inverse pattern of maximum activation time with increasing frequency was observed in MCG, contrary to TIG, and particularly in the Gamma-3 band (Fig. 3E). As such, the time delay of MCG with respect to TIG increased with frequency, peaking in the Gamma-3 band (Fig. 3D). These findings can be further extended to elaborate on the specific phenomenon observed in the Gamma-3 band in Figure 2C. To mitigate the impact of individual differences on the calculation of the time delay across participants, our statistical analysis included an examination of the time-lag metric of each frequency band relative to its alpha rhythm for each individual. Comparing the differences in the corresponding frequency bands between the two groups provided a more objective assessment of the effects of lesions on motor information transmission. Consistent with previous findings, the inhibitory effects of lesions on information flow were most prominent in the high-gamma band (Fig. 3F). Besides, ERS duration was significantly reduced in the high-frequency band as well, suggesting a weaker activation intensity within this range (Fig. 3G). These anomalies in the temporal evolution of information flow indicated that motor cortex glioma reduced the efficiency of information transfer between functional regions, and the effect exhibited significant variation across distinct frequency subbands. Collectively, the altered pathways and delayed events of information transmission indicated that the transmission of information in MCG patients is functionally distinct from that observed in patients with unaffected motor cortex.

**Glioma degraded the stability of neural decoding models.**
Given the abnormal electrophysiological activities and an atypical model for motor information transmission due to glioma, we aimed to investigate whether and how these alterations would affect the movement intention decoding from ECoG during computationally demanding tasks. For this purpose, we classified ECoG signals corresponding to movements of the thumb, index, pinky, and compound action in various frequency bands to assess the encoding capability of the motor cortex. Firstly, the assessment of recognition performance across different subbands revealed that the high-gamma band (60-140Hz) exhibited a relatively strong decoding performance compared to other subbands for both groups. However, the MCG group showed lower accuracy compared to the TIG group (45.8% vs 77.0%; $p < 0.05$), across the six frequency subbands (Fig. 4A). For the Gamma-3 band, the average movement recognition accuracy was 77.0% for TIG, with the highest and lowest individual performance of 81.0% and 71.5%, respectively, and 45.8% for MCG, with the highest and lowest individual performance of 53.5% and 39.0%, respectively (Fig. 4B).

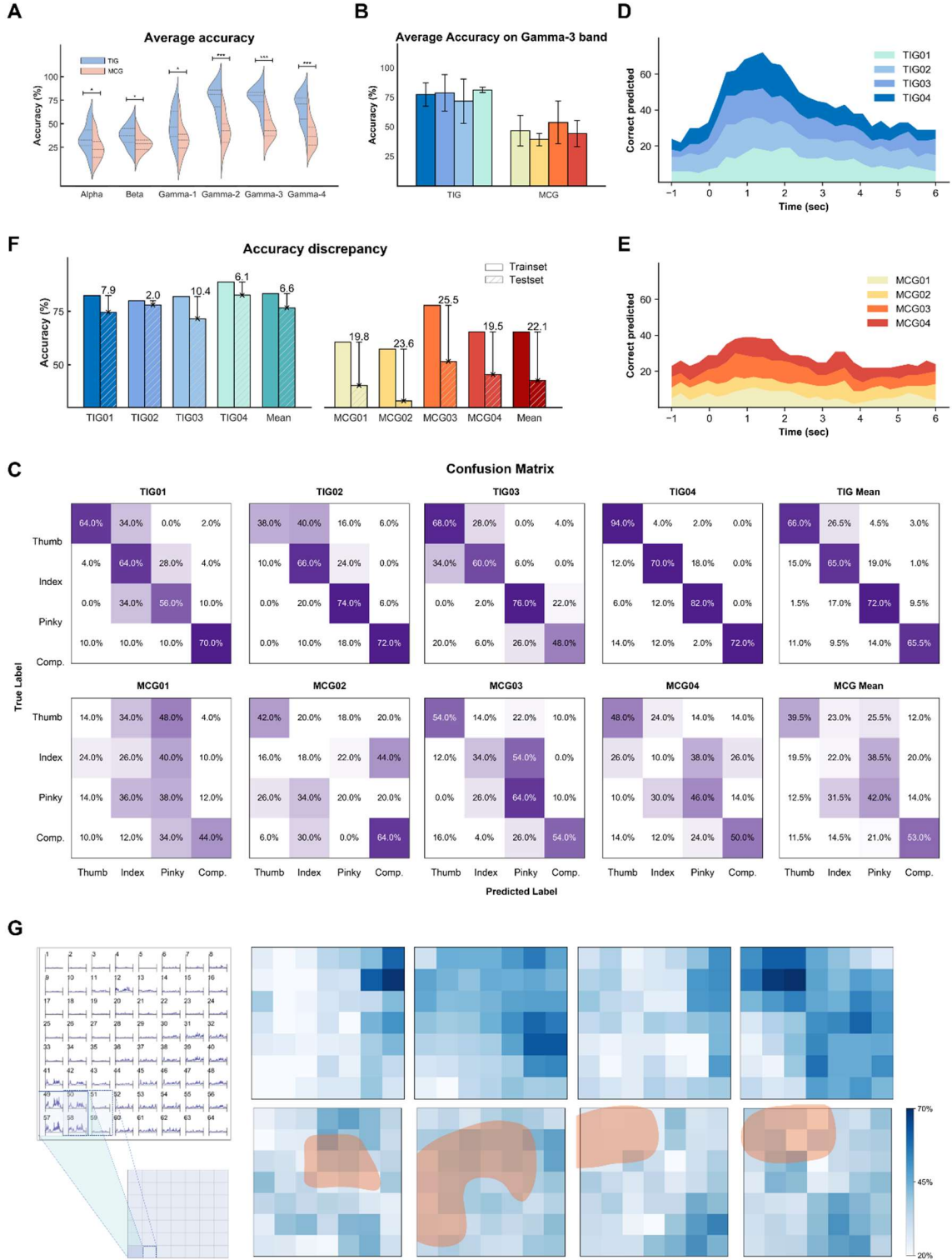

**Fig.4 Gesture decoding performance degradation in different temporal-spatial scales. (A)**, A comparative analysis of gesture recognition in different frequency bands. In the Gamma-2 and Gamma-3 bands, the decoding performance is superior for both groups. Notably, TIG group achieves a significantly higher average accuracy (nearly 80.0%), while the MCG group reaches only around 45.0%. **(B)**, The classification accuracy and standard deviation for each participant in the two groups. Within the MCG group, a noticeable decrease in the decoding accuracy of finger movements is observed for all participants in Gamma-3 band, with values ranging from roughly 39.0% to around 53.0%. **(C)**, The confusion matrix of gesture classification for each participant of both groups. In general, the mean classification accuracy of each gesture in the Gamma-3 band presents a marked decline in the MCG. **(D)-(E)**, The stacked line graphs depict the performance of movement recognition over time, showing higher accuracy during the task execution phase (1-2s). **(F)**, The accuracy discrepancy between trainset and testset can evaluate the pattern stability of signal encoding. The TIG group presents a lower discrepancy in recognition performance between the two datasets (averaging 6.6%) compared to the MCG group (average of 22.1%). It means that the information encoding stability in the MCG group has been affected. **(G)**, The local electrode partitioning method divides an 8x8 channel grid into 2x2 overlapping sliding windows with a stride of 1x1, resulting in 49 local regions. These regions are then employed for gesture recognition to evaluate the task-related local signals. The TIG group achieves optimal recognition accuracy (85.0%) at the new units, primarily concentrated in the M1 hand area. In contrast, the MCG group exhibits a deficiency in clear spatial localization regarding recognition accuracy, with comparatively lower decoding accuracy in various regions.

Compared with the gesture recognition in the TIG group, the confusion matrix for the MCG group showed smaller mean performance for all four gestures (Thumb: 66.0% vs 39.5%, Index: 65.0% vs 22.0%, Pinky: 72.0% vs 42.0%, Compound action: 65.5% vs 53.0%) (Fig. 4C). Furthermore, we analyzed the decoding performance from both temporal and spatial perspectives. Using stacked line graphs to depict the temporal properties of the decoding performance, we observed the highest fluctuations around 1-2s during the task execution phase (Fig. 4D, E). By evaluating the difference in accuracy between the training and testing sets, the stability of information coding patterns was quantified. The significant difference in accuracy between trials in the MCG group (20.1%) compared to the TIG group (6.6%) indicated that glioma impaired the stable encoding ability of the motor cortex for motor information processing (Fig. 4F). Employing the local electrode partitioning method, an 8x8 channel grid was segmented into 2x2 overlapping sliding windows with a stride of 1x1, resulting in a total of 49 local regions for assessing the task-related local signals. This allowed us to pinpoint spatial locations that exhibited optimal decoding performance. In the TIG, the highest recognition accuracy (85.0%) was achieved at the new units, primarily concentrated in the M1 hand area. However, in the MCG, the accuracy was close to chance-level (25.0%) and therefore it was not possible to distinguish among regions (Fig. 4G). However, some regions surrounding the tumor exhibited relatively high accuracy, further substantiating an atypical model of motor information transmission due to glioma.

**Discussion**

We recorded ECoG signals during awake craniotomy to study the alterations in neuronal electrical activity associated with processing hand movement from motor cortex glioma. Our findings suggest that the motor cortex around the glioma may engage in chaotic electrophysiological activity, integrate extensive functional circuits, and inefficiently encode movements during finger tasks. This information may uncover the complex pathophysiological interactions during glioma progression, and drive the implementation of neurorehabilitation approaches for patients with impaired motor cortex (19-21).

Previous studies, based on electrophysiological measurements of neural activity, suggested that neuronal oscillations within various rhythms were involved in a wide range of cognitive processes by inducing synchronization across specialized brain regions (22, 23). For example, the beta activity is commonly associated with the normal function of somatosensory and motor processes (20, 24, 25); while the gamma rhythm is linked to specific kinematic parameters of hand movements (26), motor programming (27), and visual perception (28), emphasizing its significance in cognitive activities (29). This study found that neuronal oscillations in the motor cortex infiltrated by glioma are characterized by a significant decrease in ERSP across various rhythms, with a particular reduction of ERS in the high-gamma band. Unexpectedly, the temporal relationship of ERSP curves in the Gamma-3 band between the two groups did not seem to align with the pattern observed in other gamma bands (Fig. 2C). It can be speculated that this discrepancy may be due to an excessively prolonged desynchronization duration, exceeding a fixed period. Further analyses, as depicted in Fig 3D and E, were conducted to validate this hypothesis. Despite the M1 hand area around the glioma was largely compromised, it retained the ability to coordinate neuronal oscillations following the dynamics of individual fingers, potentially explaining why participants in this study did not demonstrate noticeable neurological deficits.

The neuronal activity within the high-gamma band, specifically in the frequency range 70-170 Hz, exhibited somatotopic organization and was more spatially focused than alpha or beta oscillations (30-33). Several studies demonstrated that infiltrating glioma cells initiated disruptions in the spectral properties by affecting glutamate/GABA homeostasis, including aberrant release of the excitatory neurotransmitter glutamate, and loss of γ-aminobutyric acid (GABA) ergic inhibitory signals (34-36). The long-lasting ERD representation in the beta band may have had suppressive entrainment on the programming of movement in the high-gamma rhythm (37). In conjunction with the current results, it was speculated that the reduction in high-frequency power could be one of the mechanisms associated with the spatial dispersion of motor somatosensory localization (17).

The classic model of functional connectivity plasticity was based on a hierarchically organized model, from intratumoral and perilesional reorganization to the activation of contralateral homologs with three levels recruited sequentially (15, 38-40). When movement processing, gliomas primarily infiltrated the motor region, disrupting not only the cortical structures but also inducing abnormal electrical events in the cortex around the glioma, which impaired the interactive transmission of motor information dynamics (e.g., in the PMA - M1 circuits), and altered the spatial distribution patterns of neuronal spiking. Our analysis of cortex electrical signals evoked by motor tasks demonstrated that gliomas can reconfigure functional neuronal circuits, and activate the cortex beyond the typical cortical excitation in healthy (41, 42). This discovery may provide further insight into why gliomas, affecting the "eloquent" regions, do not recruit detectable neurological deficits in the early stage, thereby enhancing the chances of achieving total or subtotal resection.

Typically, the cortical motor representation displayed a highly organized somatotopic structure, featuring a detailed mapping of the body that corresponds with the classical homunculus model. Conversely, the presence of glioma invasion in the motor cortex leads to disorganized spatial distribution of the motor cortex and chaotic motor processing. This specific model may exhibit traits of local inefficiency, high energy consumption, and poor stability (43, 44), which contradict the small-world properties of the functional network (45-47). Additionally, the significant delay in inter-regional cortical motor information communication, particularly in the high-gamma bands, offers further evidence of their essential role in encoding finger movements and transmitting movement-related information (27, 46, 48). Hence, it is a valid assumption that the broader spatial scales in cortical neuronal activity function as a native compensatory mechanism for the limited efficiency of temporally encoded information in representing a pathologic signature of glioma infiltration.

Remarkable progress has been made in decoding motor execution and intention in recent years (49-52). However, limited attention has been paid to motor recognition from this compromised eloquent cortex. Individual finger motor decoding was found notably challenging, leading to a distinct decline in recognition accuracy, ranging from 39.0% to 53.0% in the MCG, indicating an increased instability of neural encoding properties. Also, the spatial locations with the highest decoding accuracy did not adhere to a fixed fashion based on the canonical organization of eloquent brain regions in the MCG, whereas individuals with an unaffected motor cortex achieved the highest motor recognition rates exceeding 85% using only four electrodes covering an approximate area of 1 cm2 in the M1 hand cortex. Currently, there is no consensus on the standardized location of electrodes for movement decoding in invasive brain-computer interface (BCI). This study revealed that the TIG group exhibited the highest gesture recognition performance primarily in the M1 hand knob area, providing valuable insights for electrode positioning in invasive BCI. Moreover, to accurately identify the motor functional region, a distinct strategy should be employed by covering a broader cortex with ECoG grids in cases of motor cortex infiltration by tumors, thereby opening up new avenues for studying neurorehabilitation and BCI in these specialized contexts.

Low-grade gliomas (WHO grade 1-2) progress slowly at a rate of 4-5 mm per year, and have minimal impact on quality of life, as they cause less disruption to memory, language, processing speed, and executive functions compared to glioblastoma during the pre-symptomatic stages (51). This is the clinical manifestation of the interactions between tumor progression and brain plasticity (38, 40, 53). Interestingly, we found that early signs of dysfunction caused by motor cortex gliomas were not effectively detected through cognitive assessment. Hence, stratifying functional evaluation by lesion location and imaging subtype, specifically targeting intricate finger movements, may aid in detecting subtle deficits at the onset of gliomas.

The study has some limitations. Due to the constraints of the craniotomy, the electrodes did not fully cover the primary motor cortex and other regions engaged in the motor circuit (such as the superior parietal lobule, and postcentral gyrus). Consequently, the conclusions drawn do not completely clarify the transmission of motor information from input to output. Additionally, the current study primarily focused on diffuse astrocytomas with IDH-mutant and oligodendrogliomas with IDH-mutant combined 1p/19q codeleted. While each glioma subtype may exploit a different mechanism to drive the tumor progression, future studies are required to delineate the evolution of this process and to determine whether the molecular biology of gliomas affects local neural circuitry in a more nuanced manner. Furthermore, causal relationships between disease progression and alterations in neuronal response are not yet well characterized.

**Materials and Methods**

**Participants, consent, and ethics approval.**
Each participant in the study was recruited from the Department of Neurosurgery, University of Sichuan, West China Hospital as newly diagnosed patients with diffuse glioma (Fig. S1-2). Invasive electrophysiological recordings were obtained from the motor-related cortex of eight adult glioma patients during awake craniotomies. These recordings captured the neural activity as patients executed individual finger flexion-hold-extension tasks. Inclusion criteria encompassed the following: age 18-65 years; the lesion infiltrated either the motor-related brain area or the temporo-insular lobe (Fig. S3); an electroencephalogram (EEG) was performed before the surgery, and no epileptic-like discharges were detected in electroencephalography (Neuroelectrics, Starstim32® tES-EEG systems, Austria); no previous history of psychiatric disorders, neurological illness, or drug or alcohol abuse; no severe motor dysfunction; ability to cooperate and complete motor task during awake surgery; the initial magnetic resonance imaging (MRI) diagnosed diffuse glioma, and was finally confirmed by the 2021 World Health Organization (WHO) classification(18). Exclusion criteria included neuropsychiatric disorders or comorbidities of the central nervous system; previously diagnosed with glioma and underwent stereotactic radiotherapy, chemotherapy, or surgery; pathological diagnosis rules out gliomas.

All participants provided written informed consent to participate in this study, which was approved by the institutional review board at the University of Sichuan, West China Hospital (No. 2023-1231), and performed following the Declaration of Helsinki.

**Awake craniotomy and cortical stimulation mapping.**
During awake surgeries, all participants underwent intraoperative mapping using neuro-navigation devices (BrainLab AG, Munich, Germany) to determine the location and extent of the lesion. Neuronal electrical stimulation was administered using a bipolar stimulator (Nicolet Cortical Stimulator, Natus Medical Incorporated, Middleton, WI, USA) at a 5mm interval, 60Hz frequency, and 1 ms pulse width. Stimulation currents ranged from 1.0 to 5.0mA with a 2-5 second duration(3). Cortical direct stimulation (DCS) was performed for the precise localization of the central precentral gyrus, central postcentral gyrus, and Broca's area. A positive response was recorded upon induction of movement in the contralateral limbs, and the corresponding areas were labeled and documented as DCS-positive electrodes. Images of the brain surface were captured with a DCS-positive site marker of 3mm radius.

**Finger movement task paradigm.**
Each patient received a training session 24 hours before the craniotomy to ensure familiarity with the flexion-holding-extension of individual fingers task. A 15-inch laptop computer running a custom Python script (Version 3.7) with a 60Hz refresh rate was positioned 30 cm away from each patient. Upon awakening from anesthesia, patients were instructed to perform randomly assigned tasks in response to cues provided by the video on the screen.

During the electrocorticography (ECoG) recordings, the experiment consisted of five sequential runs, each lasting about 49 seconds, interspersed with 3-second epochs of rest between runs, resulting in a total duration of approximately 5 minutes. In detail, a 2-minute resting state of the ECoG signal was collected before the start of the motor task. During a trial, patients were instructed to execute designed task paradigms, holding finger flexion for about 4 seconds, followed by a rest period of 2 seconds. To minimize noise in the operating room, strict measures were enforced, including instructing all personnel to refrain from speaking, muting or temporarily shutting down telephones, machinery alarms, surgical suction, and other nonessential equipment (Fig. 1B-C).

**Electrodes localization and signal acquisition.**
The platinum electrode arrays (HKHS Healthcare Co., Ltd., Beijing, China), embedded within a plastic sheet, had a contact surface diameter of 2mm and were spaced 5mm apart from each other, typically configured as 64 or 128 electrodes. The signals were recorded using the Synamps2 system (g.tec medical engineering, GmbH, Schiedlberg, Austria) with an analog bandwidth filter of direct-current systems to 200 Hz and a notch filter of 50 Hz, or the NeuSen H (Neuracle Technology Co., Ltd., Changzhou, China), which amplified and sampled the data at a minimum of 1 kHz and a maximum of 2 kHz of sampling rate. Channels with excessive amounts of noise were excluded from the analysis. Needle electrodes at the ipsilateral and contralateral scalp were set as ground and reference, respectively.

By matching the stereotactic neuro navigation, intraoperative photographs of the cortex with the overlying grid, and anatomical landmarks on the MRI, each electrode was registered to the MRI T2-weighted-fluid-attenuated inversion recovery (FLAIR) image or T1 post gadolinium using the MRICro software (http://www.mccauslandcenter.sc.edu/mricro/). Moreover, these electrodes were classified as "glioma" or "normal-appearing" by a trained neurosurgeon. In detail, glioma regions were defined based on two criteria previously established in the literature(17): when the electrode site corresponding to the cerebral cortex showed aberrant signals in the axial, sagittal, and coronal images of the FLAIR image, it was labeled as a glioma-area electrode. A completely normal signal in all three dimensions was defined as a normal cortex electrode. These classifications were reviewed by the study principal investigator (YY, QM) to reach a consensus, while keeping the classification results blinded to the electrophysiological specialist during signal processing (Fig. S4).

Ultimately, the coordinates of these electrodes, obtained from the surgical photographs, were mapped onto a standardized brain model—the Montreal Neurological Institute (MNI) template space 0.8 mm isotropic volumetric map—using Brainstorm (https://neuroimage.usc.edu/brainstorm/), for each participant, to facilitate group comparisons (Fig. 1D). A kernel density estimation (KDE) was utilized to evaluate the distribution of electrodes positions corresponding to the tumor in MCG group (R Foundation, Vienna, Austria, version 4.1.2) (Fig. S5).

**ECoG signal preprocessing.**
The ECoG signal preprocessing involved several key steps. Initially, a notch filter was utilized to eliminate power line interference at 48-52 Hz and its harmonics at 98-102 Hz. Following this, a FIR bandpass was employed to extract information within the 4-140 Hz frequency range. The data was then time-locked by aligning it with the start of the task and designated as the reference point (zero). A time window from -1 to 6 seconds was selected, encompassing a 1-second baseline period before task initiation and a 6-second period covering task execution and maintenance. Any channels or segments displaying poor performance were identified and preprocessed using global mean interpolation to ensure overall signal quality. The entire dataset underwent common average re-referencing to eliminate common-mode noise artifacts. The channel displaying the most significant power modification before and after the cue illustration in the M1 hand area was identified as the targeted electrode, as illustrated in Fig. 2A. Subsequently, the signals were divided into six frequency sub bands based on empirical definitions: Alpha (8-13 Hz), Beta (13-30 Hz), Gamma-1 (30-60 Hz), Gamma-2 (60-90 Hz), Gamma-3 (90-120 Hz), and Gamma-4 (120-140 Hz). To explore the impact of glioma on the motor cortex, ERSP and power were computed across each sub band.

**The temporal-frequency features of neuronal activity across various frequency bands.**
Through the analysis of ERSP across various frequency bands, we aim to comprehensively understand the impact of glioma on neuronal oscillations in the motor cortex. To perform a detailed time-frequency analysis for each individual, we utilized the signal segment spanning from -1 to 6 seconds during the finger movement task, and computed ERSP for each specific finger action (Eq. 1-2).

$$ERS^{i,j} = \frac{1}{n}\sum_{k=1}^{n} |F^{i,j}(f,t)|^2 \qquad (1)$$

$$ERSP^{i,j} = 10 * \log_{10} \frac{ERS^{i,j}_{task}}{ERS^{i,j}_{base}} \qquad (2)$$

where n represents the number of trials conducted for a specific task, $i$ denotes the participant number, and $j$ refers to the task number. $F(f,t)$ denotes the spectral estimate at frequency $f$ and time $t$, $ERS^{i,j}_{task}$ and $ERS^{i,j}_{base}$ indicate the ERS features during the task state and baseline state, respectively. To compare the time-frequency features within each group, we calculated $ERSP^{i,j}$ and averaged the ERSP values across the tasks for each finger (Eq. 3).

$$ERSP = \frac{1}{p}\sum_{i=1}^{p} \frac{1}{q}\sum_{j=1}^{q} ERSP^{i,j} \qquad (3)$$

The variable $p$ corresponds to the number of participants in the group, while $q$ denotes the number of tasks for each participant. The result provides an illustration of the time-frequency features at frequency $f$ and time $t$, as depicted in Fig. 2B. Subsequently, through the computation of the average ERSP for various temporal points within each frequency band, we generated

ERSP curves across different bands between the two groups. Time points demonstrating significant differences (p < 0.05) are denoted in bold. We then calculated the Pearson correlation coefficient (PCC) to assess the similarity between the curves (Fig. 2C). Additionally, to analyze the temporal dynamics of the ERSP fluctuation, we applied Welch's ANOVA statistical analysis to identify discrepancies between the two curves in distinct frequency bands (Fig. 2D). Furthermore, we computed the average power of each frequency band during the execution period ([0, 3] seconds), which could be described as Eq. 4, and employed a Wilcoxon rank-sum test to compare the power between the two groups within each frequency band (Fig. 2E).

$$Power = 10 * \log_{10} \frac{1}{T} \sum_{t=1}^{T} |f(t)|^2 \qquad (4)$$

Here, $f(t)$ signifies the signal of a specific frequency band, and $T$ represents the total number of sampling points during the execution period.

**The temporal-spatial features of motor information transmission across various frequency bands.**

Based on the previously mentioned research findings, we proceed to investigate the impact of MCG on the transmission of motion information in the high-frequency range. Initially, the phase-locking value (PLV) was calculated for various channels within the high-gamma frequency range to assess information transmission efficiency, which is formulated as Eq. 5.

$$PLV = \frac{1}{N} \left| \sum_{i=1}^{N} e^{i(\varphi_1(i) - \varphi_2(i))} \right| \qquad (5)$$

Here, $T$ signifies the number of data points, while $\varphi_1(i)$ and $\varphi_2(i)$ represent the instantaneous phase values of two channels at time point $i$. Additionally, $e$ denotes Euler's number. In addition, we assessed the most significant 50 sets of phase desynchronization involving channels and visualized their spatial distribution to demonstrate their spatial distribution (Fig. 4C).

The instantaneous phase was derived from a narrow frequency range, specifically from three sub bands (70-80 Hz, 80-90 Hz, and 90-100 Hz) within the high-gamma band, and the mean PLV values of these sub bands were computed to estimate the PLV results of the specific frequency band. PLV values range from 0 to 1, with a higher value indicating a greater phase synchronization between the two channels. During a task, the presence of lock-to-phase relationships between different channels suggests that the activation order of these channels is correlated, indicating the existence of information transfer between them. However, the direction of this information transfer is not specified. To visualize these phenomena, we rearranged all channels based on their spatial relationships, focusing on the tumor area, as shown in Fig. 3A and B.

Through computing the cross-correlation (CC) for both groups during the task execution phase (0 - 3 seconds), we can identify the time point with the strongest correlation and discern the disparity in the timing of reaching maximum activation within specific brain regions during this phase (Fig.

3D). Furthermore, a comparison of the maximum activation time across various frequency bands between the two groups in Fig.3E. This process can be described as:

$$ccor_{xy}(n) = \sum_{m=0}^{M-n} x(m) * y(m+n) \tag{6}$$

$$T = \arg\max_{n} ccor_{xy}(n) \tag{7}$$

where $x$ and $y$ denote two signals, $x(m)$ represent the value of signal $x$ at time $m$, $M$ is the total length of the signal, $ccor_{xy}(n)$ is the cross-correlation value between $x$ and $y$ with a delay of $n$, and $T$ signifies the calculated delay time for information transfer.

Considering inter-individual variability in the task process, we separately computed the ratio of time lag in each frequency band to that in the alpha band for each group. Subsequently, we analyzed the difference in the corresponding frequency bands between the two groups (Fig.3F). In addition, we quantified the activation intensity and duration of the high-gamma band by calculating the proportion of time that the ERS phenomenon occurred in the high-gamma frequency range (60-140 Hz) between 0 to 3 seconds (Fig.3G). These indicators provide valuable information into the efficiency of motor circuit interaction between various regions.

Apart from measuring the temporal dimension, by quantifying the power of electrodes positioned in various spatial locations, we evaluated the extent to which these electrodes corresponded to cortex regions engaged in the transmission of motor information. Specifically, during the task execution phase (0 - 3s), we computed the power of each electrode, normalized the values within the range of minimum and maximum range, and then mapped all electrodes onto a template brain based on their intraoperative location (Fig.3H). By mapping the electrodes onto the template, we visually represented the spatial distribution of electrical activity across different brain regions during task processing, where the size of each circle correlates positively with its normalized value.

**The temporal-spatial features of gesture recognition performance across various frequency bands.**
We employed gesture recognition performance of different frequency bands to assess the programming ability of the motor cortex in representing movements. Firstly, we extracted the ECoG signals corresponding to the four movement of the fingers, i.e., thumb, index, pinky and a compound action, across six frequency sub-bands. These signals from each trial were divided into multiple samples using a 1-second window with a 0.2-second stride, and separate action classification was conducted for each frequency band. Then, we computed the power of each action at every time point for each participant, and principal component analysis (PCA) was employed to extract significant features by decreasing the dimension of every frequency band. We then employed linear discriminant analysis (LDA) to classify these features. Finally, to determine the average accuracy for each participant across various frequency bands, leave-one-out cross-validation experiments were conducted at the session level.

The classification accuracy for both groups was computed, using features extracted from each frequency band. The average classification performance across various frequency bands for both groups was depicted in Fig.4A, with a specific emphasis on the best classification performance in the Gamma-3 band for each participant (Fig.4B). Among these, the confusion matrix for various gestures in the best frequency band, Gamma-3, was analyzed and presented (Fig.4C).

Moreover, we examined the decoding accuracy of ECoG signals during various task phases. The model was initially trained using signals from the action execution stage ([0, 3] s) on the trainset, followed by evaluating its effectiveness on the entire testset ([-1, 6] seconds). An average gesture recognition accuracy at each time point was calculated (Fig.4D-E), providing an understanding of the temporal dynamics of task relevance and the stability of neural encoding models. We also computed the differences in accuracy between the training and testing sets for each subject to measure the stability of the information encoding patterns. A smaller difference indicates that the training and testing sets have more similar distributions, thus indicating a more stable information encoding pattern (Fig.4F).

Additionally, we applied the local electrode partitioning method to divide an 8x8 channel grid into 2x2 overlapping sliding windows with a stride of 1x1, resulting in a total of 49 local regions to evaluate the decoding performance (Fig.4G). Our analysis centered on the channels within the designated unit to conduct classification tasks. We then investigated the average action recognition features of the newly implemented electrode unit, the higher of which revealed the spatial regions with more stable signal patterns and stronger task relevance.

**Acknowledgments**

**Funding**
Sichuan Science and Technology Program (Grant no. 2023YFG0127)
1.3.5 project for Disciplines of 1435 Excellence Grant from West China Hospital under Grant ZYYC22001
National Natural Science Foundation of China under Grant no. 82072773
Fundamental Research Funds for the Central Universities under Grant YJ202373

**Data and materials availability**
The data supporting the findings of this study are not publicly available as they may compromise the privacy and consent of research participants. Anyone seeking access to the research-related data for validation or further study is required to request permission from the corresponding authors. Upon approval, the data will be made available to the authorized parties.

**References**

1. N. A. Mohile *et al.*, Therapy for Diffuse Astrocytic and Oligodendroglial Tumors in Adults: ASCO-SNO Guideline. *J Clin Oncol* **40**, 403-426 (2022).
2. T. Saito *et al.*, Awake craniotomy with transcortical motor evoked potential monitoring for resection of gliomas within or close to motor-related areas: validation of utility for predicting motor function. *J Neurosurg* **136**, 1052-1061 (2022).
3. A. J. Gogos, J. S. Young, R. A. Morshed, S. L. Hervey-Jumper, M. S. Berger, Awake glioma surgery: technical evolution and nuances. *J Neurooncol* **147**, 515-524 (2020).
4. J. Tuominen, S. Yrjana, A. Ukkonen, J. Koivukangas, Awake craniotomy may further improve neurological outcome of intraoperative MRI-guided brain tumor surgery. *Acta Neurochir (Wien)* **155**, 1805-1812 (2013).
5. S. Fang, Y. Li, Y. Wang, Z. Zhang, T. Jiang, Awake craniotomy for gliomas involving motor-related areas: classification and function recovery. *J Neurooncol* **148**, 317-325 (2020).
6. S. L. Hervey-Jumper *et al.*, Awake craniotomy to maximize glioma resection: methods and technical nuances over a 27-year period. *J Neurosurg* **123**, 325-339 (2015).
7. P. McColgan, J. Joubert, S. J. Tabrizi, G. Rees, The human motor cortex microcircuit: insights for neurodegenerative disease. *Nat Rev Neurosci* **21**, 401-415 (2020).
8. S. Ng *et al.*, Intraoperative functional remapping unveils evolving patterns of cortical plasticity. *Brain* **146**, 3088-3100 (2023).
9. H. S. Venkatesh *et al.*, Electrical and synaptic integration of glioma into neural circuits. *Nature* **573**, 539-545 (2019).

10. S. Krishna *et al.*, Glioblastoma remodelling of human neural circuits decreases survival. *Nature* **617**, 599-607 (2023).
11. T. Numan *et al.*, Regional healthy brain activity, glioma occurrence and symptomatology. *Brain* **145**, 3654-3665 (2022).
12. A. D. Friederici, N. Chomsky, R. C. Berwick, A. Moro, J. J. Bolhuis, Language, mind and brain. *Nat Hum Behav* **1**, 713-722 (2017).
13. Y. Zhu *et al.*, Distinct spatiotemporal patterns of syntactic and semantic processing in human inferior frontal gyrus. *Nat Hum Behav* **6**, 1104-1111 (2022).
14. G. K. Anumanchipalli, J. Chartier, E. F. Chang, Speech synthesis from neural decoding of spoken sentences. *Nature* **568**, 493-498 (2019).
15. F. Rech, H. Duffau, Beyond Avoiding Hemiplegia after Glioma Surgery: The Need to Map Complex Movement in Awake Patient to Preserve Conation. *Cancers (Basel)* **15**, (2023).
16. M. Zetterling, K. Elf, R. Semnic, F. Latini, E. R. Engstrom, Time course of neurological deficits after surgery for primary brain tumours. *Acta Neurochir (Wien)* **162**, 3005-3018 (2020).
17. A. A. Aabedi *et al.*, Functional alterations in cortical processing of speech in glioma-infiltrated cortex. *Proc Natl Acad Sci U S A* **118**, (2021).
18. D. N. Louis *et al.*, The 2021 WHO Classification of Tumors of the Central Nervous System: a summary. *Neuro Oncol* **23**, 1231-1251 (2021).
19. J. P. Savarraj, K. C. Kelly, M. A. DeCoster, Early glioma is associated with abnormal electrical events in cortical cultures. *Med Biol Eng Comput* **57**, 1645-1656 (2019).
20. P. J. Uhlhaas, C. Haenschel, D. Nikolic, W. Singer, The role of oscillations and synchrony in cortical networks and their putative relevance for the pathophysiology of schizophrenia. *Schizophr Bull* **34**, 927-943 (2008).
21. V. Belgers *et al.*, Postoperative oscillatory brain activity as an add-on prognostic marker in diffuse glioma. *J Neurooncol* **147**, 49-58 (2020).
22. A. K. Engel, C. Gerloff, C. C. Hilgetag, G. Nolte, Intrinsic coupling modes: multiscale interactions in ongoing brain activity. *Neuron* **80**, 867-886 (2013).
23. T. Numan *et al.*, Non-invasively measured brain activity and radiological progression in diffuse glioma. *Sci Rep* **11**, 18990 (2021).
24. M. Illman, K. Laaksonen, V. Jousmaki, N. Forss, H. Piitulainen, Reproducibility of Rolandic beta rhythm modulation in MEG and EEG. *J Neurophysiol* **127**, 559-570 (2022).
25. J. L. Ulloa, The Control of Movements via Motor Gamma Oscillations. *Front Hum Neurosci* **15**, 787157 (2021).
26. T. Pistohl, T. Ball, A. Schulze-Bonhage, A. Aertsen, C. Mehring, Prediction of arm movement trajectories from ECoG-recordings in humans. *J Neurosci Methods* **167**, 105-114 (2008).
27. N. E. Crone, D. L. Miglioretti, B. Gordon, R. P. Lesser, Functional mapping of human sensorimotor cortex with electrocorticographic spectral analysis. II. Event-related synchronization in the gamma band. *Brain* **121 ( Pt 12)**, 2301-2315 (1998).
28. D. Hermes, K. J. Miller, B. A. Wandell, J. Winawer, Gamma oscillations in visual cortex: the stimulus matters. *Trends Cogn Sci* **19**, 57-58 (2015).
29. T. Jiang *et al.*, Characterization and Decoding the Spatial Patterns of Hand Extension/Flexion using High-Density ECoG. *IEEE Trans Neural Syst Rehabil Eng* **25**, 370-379 (2017).
30. M. T. Kucewicz *et al.*, High frequency oscillations are associated with cognitive processing in human recognition memory. *Brain* **137**, 2231-2244 (2014).
31. F. Artoni *et al.*, High gamma response tracks different syntactic structures in homophonous phrases. *Sci Rep* **10**, 7537 (2020).
32. S. Raspopovic *et al.*, Neural signal recording and processing in somatic neuroprosthetic applications. A review. *J Neurosci Methods* **337**, 108653 (2020).
33. Y. Dong *et al.*, Neural Decoding for Intracortical Brain-Computer Interfaces. *Cyborg Bionic Syst* **4**, 0044 (2023).

34. M. K. Montgomery *et al.*, Glioma-Induced Alterations in Neuronal Activity and Neurovascular Coupling during Disease Progression. *Cell Rep* **31**, 107500 (2020).
35. S. C. Buckingham *et al.*, Glutamate release by primary brain tumors induces epileptic activity. *Nat Med* **17**, 1269-1274 (2011).
36. S. L. Campbell *et al.*, GABAergic disinhibition and impaired KCC2 cotransporter activity underlie tumor-associated epilepsy. *Glia* **63**, 23-36 (2015).
37. K. J. Miller *et al.*, Human motor cortical activity is selectively phase-entrained on underlying rhythms. *PLoS Comput Biol* **8**, e1002655 (2012).
38. H. Duffau, Lessons from brain mapping in surgery for low-grade glioma: insights into associations between tumour and brain plasticity. *Lancet Neurol* **4**, 476-486 (2005).
39. H. Duffau, The huge plastic potential of adult brain and the role of connectomics: new insights provided by serial mappings in glioma surgery. *Cortex* **58**, 325-337 (2014).
40. H. Duffau, Diffuse low-grade glioma, oncological outcome and quality of life: a surgical perspective. *Curr Opin Oncol* **30**, 383-389 (2018).
41. R. H. Sanders, D. J. Levitin, Towards an Understanding of Control of Complex Rhythmical "Wavelike" Coordination in Humans. *Brain Sci* **10**,  (2020).
42. A. Battaglia-Mayer, R. Caminiti, Corticocortical Systems Underlying High-Order Motor Control. *J Neurosci* **39**, 4404-4421 (2019).
43. L. Pasquini *et al.*, Brain Functional Connectivity in Low- and High-Grade Gliomas: Differences in Network Dynamics Associated with Tumor Grade and Location. *Cancers (Basel)* **14**,  (2022).
44. L. Douw, L. C. Breedt, M. L. M. Zimmermann, Cancer meets neuroscience: the association between glioma occurrence and intrinsic brain features. *Brain* **146**, 803-805 (2023).
45. F. Vecchio *et al.*, Human brain networks: a graph theoretical analysis of cortical connectivity normative database from EEG data in healthy elderly subjects. *Geroscience* **42**, 575-584 (2020).
46. I. Indovina, J. N. Sanes, On somatotopic representation centers for finger movements in human primary motor cortex and supplementary motor area. *Neuroimage* **13**, 1027-1034 (2001).
47. H. Xu *et al.*, Reduced efficiency of functional brain network underlying intellectual decline in patients with low-grade glioma. *Neurosci Lett* **543**, 27-31 (2013).
48. E. L. Rich, J. D. Wallis, Spatiotemporal dynamics of information encoding revealed in orbitofrontal high-gamma. *Nat Commun* **8**, 1139 (2017).
49. K. J. Miller, D. Hermes, N. P. Staff, The current state of electrocorticography-based brain-computer interfaces. *Neurosurg Focus* **49**, E2 (2020).
50. M. G. Bleichner *et al.*, Give me a sign: decoding four complex hand gestures based on high-density ECoG. *Brain Struct Funct* **221**, 203-216 (2016).
51. C. E. Bouton *et al.*, Restoring cortical control of functional movement in a human with quadriplegia. *Nature* **533**, 247-250 (2016).
52. L. R. Hochberg *et al.*, Reach and grasp by people with tetraplegia using a neurally controlled robotic arm. *Nature* **485**, 372-375 (2012).
53. H. Duffau, New concepts in surgery of WHO grade II gliomas: functional brain mapping, connectionism and plasticity--a review. *J Neurooncol* **79**, 77-115 (2006).